\documentclass[12pt]{article}
 \usepackage[ruled,vlined,linesnumbered]{algorithm2e}
\usepackage{booktabs,amsfonts,color}
\usepackage{graphicx}
\usepackage{amsmath,bm}
\usepackage{comment}
\usepackage{multirow}
\usepackage{setspace}
\usepackage{float}
\usepackage{color}
\usepackage{indentfirst}
\usepackage{epstopdf}
\usepackage{epsfig}

\usepackage{amssymb}
 
\usepackage{lscape}
\usepackage{bm}
\usepackage{natbib}
\usepackage{tikz}
\usepackage{rotating}
\usepackage{multirow}
\usepackage{hyperref}

 \usepackage{subfig} 

\usepackage{booktabs,amsfonts,color}
 
\newcommand{\be}{\begin{equation}}
\newcommand{\ee}{\end{equation}}
\newcommand{\bea}{\begin{eqnarray}}
\newcommand{\eea}{\end{eqnarray}}
\newcommand{\beaa}{\begin{eqnarray*}}
\newcommand{\eeaa}{\end{eqnarray*}}

\def\X{\mathbf{X}}
\def\1{\mathbf{1}}

\def\Xi{\bm{\xi}}

 \def\O{\mathcal{O}} 
 
\def\so{{\scriptscriptstyle{\mathcal{O}}}}

\def\D{\mathcal{C}}

\definecolor{darkgreen}{rgb}{0, .6, 0}

\long\def\symbolfootnote[#1]#2{\begingroup%
\def\thefootnote{\fnsymbol{footnote}}\footnote[#1]{#2}\endgroup}
\date{}
\begin{document}
 \def\spacingset#1{\renewcommand{\baselinestretch}%
{#1}\small\normalsize} \spacingset{1}

 \title{Two edge-count tests and relevance analysis in k high-dimensional samples }

\author{Xiaoping Shi\thanks{Irving K. Barber School of Arts and Sciences,  University of British Columbia,  Kelowna,  BC V1V 1V7, Canada.
E-mail: xiaoping.shi@ubc.ca}}
  \date{}
  \maketitle
 
\begin{abstract}

 For the task of relevance analysis, the conventional Tukey's test may be applied   to the set of all pairwise comparisons. However, there were few studies that discuss both nonparametric k-sample comparisons and relevance analysis in high dimensions.
  Our aim is to  capture the degree of relevance between combined samples and provide additional insights and advantages in high-dimensional k-sample comparisons.
  Our solution is to extend a graph-based two-sample comparison and investigate its availability for large and unequal sample sizes.
We propose two distribution-free test statistics  based on between-sample edge counts and      measure the degree of relevance by standardized counts. 
 The asymptotic permutation null distributions of the proposed statistics are derived, and the power gain is proved when the sample sizes are smaller than the square root of the dimension.  
   We also discuss different edge costs in the  graph to compare  the parameters of the distributions.
 Simulation comparisons and real data analysis of tumors and images  further convince the value of our proposed method.
 Software implementing the relevance analysis  is available in the R package Relevance.

\end{abstract}


\allowdisplaybreaks

\newpage
\spacingset{1.45} 

\section{Introduction}

We are now in an era of data explosion. Due to transparency and the right to data, improved technology for accessing data, and increased data storage capacity, vast amounts of high-dimensional data, including thousands of variables (features or attributes) and hundreds of instances
are being entered into commercial, humanities, medical, scientific, and government databases around the world.  
Here are some examples of high-dimensional data.

{\bf 
Microarray data:}  
 The gene expression microarray technology, which can measure tens of thousands of gene expression levels in a single experiment \citep{SSB96}, have proven to be a very mature technology. With the illumina HiSeq platform, 20,531 gene expression levels were extracted from 801 patients   to compare five different types of tumors for effective therapies \citep{We13}. These data are available from
\url{https://archive.ics.uci.edu/ml/datasets/gene+expression+cancer+RNA-Seq}. 

{\bf Proteomic data:} Aptamers, single-stranded oligonucleotides, are    advanced tools for measuring plasma proteins \citep{ZR17}. Through 
venipuncture, \cite{Le19} measured 2,925 plasma proteins to capture aspects of aging in different cell types and tissues. They compared dozens of instances with thousands of protein levels by applying the so-called sliding window analysis. Some of the high-dimensional data can be obtained from the R package DEswan at \url{https://lehallib.github.io/DEswan/}.

{\bf Functional magnetic resonance imaging (fMRI) data:}
The physiological monitoring system (Model 1030, Small Animal Instruments, Stony Brook, USA) allows continuous measurement of eletrocardiogram (ECG) and motion-sensitive respiration signals. By inserting 30G needle electrodes, \cite{JSK19} collected blood oxygen level-dependent (BOLD) fMRIs from seven adult males under 9.4 T and 15.2 T magnetic resonance (MR) conditions. Their aim was to compare the functional detection of BOLD responses to certain stimuli by these two MR conditions under different experimental periods: pre-stimulus (40s), stimulus (20s) and post-stimulus groups (60s).
The size of the collected fMRI images  is 228$\times$160$\times$132$\times$120.


\subsection{Problems}

In traditional data collection, limited by technology, we usually have many observations and several variables. In order to compare whether the linear combination of the created mean vectors differs from one sample to another, we can perform a classical multivariate analysis of variance (MANOVA) with two important assumptions: multivariate normality and homogeneity of variance.

Nowadays, we collect data on genes, proteins or images. The dimensionality of each observation is in the thousands, while there are only tens or hundreds of instances available for study. MANOVA simply cannot cope with this high dimensionality, nor can it compare sample covariances. As an example of a proteomic comparison, we have four samples: males after age $t$ (MA), females after age $t$ (FA), males before age $t$ (MB), and females before age $t$ (FB).  We assume that the mean vectors of the protein samples are $\bm{\mu}_{\text{MA}}$,  
$\bm{\mu}_{\text{FA}}$, $\bm{\mu}_{\text{MB}}$ and $\bm{\mu}_{\text{FB}}$, respectively.  The \textit{linear} combination of mean vectors,  
$(\bm{\mu}_{\text{MA}}-\bm{\mu}_{\text{MB}})+(\bm{\mu}_{\text{FA}}-\bm{\mu}_{\text{FB}})$, represents the difference between two different age samples before and after $t$. However, in practical applications, there may be a negative correlation between males and females. Therefore, the   weighted sum of norms, $w_1||\bm{\mu}_{\text{MA}}-\bm{\mu}_{\text{MB}}||+w_2||\bm{\mu}_{\text{FA}}-\bm{\mu}_{\text{FB}}||$, should be more reasonable. But it is not trivial to determine the weights $w_1$ and $w_2$, which could be related to sample sizes and covariances. Developing   nonparametric methods may be a good way to compare sample distributions and perform relevance analysis simultaneously in high dimensions.

\subsection{Literature review}

In a two-sample comparison, the classical Hotelling's $T^2$ statistic is $(\bar{\bm{X}}_1-\bar{\bm{X}}_2)^\top S^{-1}(\bar{\bm{X}}_1-\bar{\bm{X}}_2)$, where $\bar{\bm{X}}_1$ and $\bar{\bm{X}}_2$ are the two sample mean vectors and $S$ is the pooled sample covariance. If the dimensionality is larger than the sample size, the inverse of $S$ may not exist, so Hotelling's $T^2$ statistic does not work.
An important progress was initially made by \cite{BS96} by introducing $(\bar{\bm{X}}_1-\bar{\bm{X}}_2)^\top (\bar{\bm{X}}_1-\bar{\bm{X}}_2)$. They derived its asymptotic distribution under normality and homogeneous covariance. \cite{CQ10} extended this by considering an U-statistic.   Cai et al. (2014) proposed an extreme statistic by comparing each component. \cite{XLWP16} considered a $\gamma$-norm ($1\leq\gamma<\infty$) of $\bar{\bm{X}}_1-\bar{\bm{X}}_2$. The $\gamma$-norm of $d$-dimensional vector $\bm{X}$ is defined by $||\bm{X}||_\gamma=(\sum_{q=1}^d|X_q|^\gamma)^{1/\gamma}$, where $X_q$ is the $q$th component of this vector. 
Thereafter, $||\bar{\bm{X}}_1-\bar{\bm{X}}_2||^2_2=(\bar{\bm{X}}_1-\bar{\bm{X}}_2)^\top (\bar{\bm{X}}_1-\bar{\bm{X}}_2)$. The introduction of the $\gamma-$norm 
 paved the way for the graph method; see the cost \eqref{GN}. \cite{HXWP21} further proposed a family of U-statistics as an unbiased estimate of the $\gamma$-norm.

Along the development of U statistics, \cite{SK13} compared the means of several samples with a common covariance matrix, while \cite{HBWW17} relaxed it to unequal covariance matrices. These research advances indicate the feasibility of extending the two-sample comparison to the k-sample comparison in terms of the combination of the U-statistic and the $\gamma$-norm.
It should not be overlooked that there is also a graph method based on graphical optimization that treats data of arbitrary dimension as points and minimizes the total costs under some constraints such as tree and path.

Denote a set of points (nodes or vertices) by $\mathcal{V}=\{v_1,\ldots,v_N\}$. Let $\mathcal{G}$ be a connected and costed graph with a set of edges $\mathcal{E}(\mathcal{G})$. We assign $\gamma$-norm to the cost of each edge. An information graph can have some features sensitive to alternatives. Currently, two kinds of optimized graphs are studied: minimum spanning tree (MST) and shortest Hamiltonian path (SHP).
Both tests based on MST \citep{FR79} and SHP \citep{BMG14}  can be  generalized  from the Wald-Wolfowitz run  test \citep{WW40}. However, the MST-based test that counted the between-sample edges can result in the power loss \citep{CZ13, CCS18}.
Some improvements in  power gain have been made through counting the within-sample edges \citep{CF17, ZC21}.
For other developments in nonparametric tests, see the recent work of \citep{MW20}.

 

\subsection{Contributions}
 
 Our main contributions include:
\begin{itemize}
\item[1.]
It is still unknown whether the two-sample edge counts can be used to solve the k-sample comparison problem. In addition, the power gain at large sample sizes is uncertain. We convince the applicability  and power gain of this counting by requiring the sample sizes to be smaller than the square root of the dimension. This allows the sample sizes to grow with the dimension.

\item[2.] We propose two distribution-free tests based on the weighted sum of between-sample counts and minimum of standardized between-sample counts,  and derive  asymptotic permutation null distributions.
 The degree of relevance between combined samples can be  measured   by the standardized between-combined-sample counts.

\item[3.] We also discuss how to accurately assess the  degree relevance by choosing appropriate edge costs.  We call our technique \texttt{relevance}, short for \underline{rel}ated \underline{e}ntire \underline{v}ariety \underline{an}alysis of \underline{c}ounts of \underline{e}dges; it is implemented in the \textbf{R} package \textbf{Relevance}. The simulation study demonstrates the advantage of our test for the k-sample comparison problem. Two real data analyses further convince the value of relevance analysis.

\end{itemize}

 For convenience, we make these notations. Write the vectors $\X_t=(X_{t,1},\ldots,X_{t,d})^\top$, the difference vectors 
 $\dot{\X}_t=(X_{t,2}-X_{t,1},\ldots,X_{t,d}-X_{t,d-1})^\top$, and the partial sums $X_{t\bullet}=\sum_{j=1}^dX_{t,j}$ for $1\leq t\leq N$.
Denote  $k$ samples indexed by $g_t$, where for any $t$, there exists only one $G_\ell\subseteq\mathbb N$ with $1\leq \ell\leq k$ such that $g_t\in G_\ell$. Denote each size as $n_\ell=|G_\ell|$ or $\#\{G_\ell\}$ and total size as   $N=\sum_{\ell=1}^kn_\ell$. Write $\bm{\theta}(F_{G_\ell})$  as the parameter vector of the common distribution $F$ of the sample $G_\ell$. The multivariate normal distribution is denoted as $N(\bm{0}, \bm{\Sigma})$   with a   mean vector $\bm{0}=(0,0,\ldots,0)^\top$ and a covariance matrix $\bm{\Sigma}$. In particular, $\mathbb{I}$ is an identity matrix. We denote the  $N!$ distinct paths as $\mathcal{P}_{\text{all}}=\{\text{all~distinct~paths~}v_{q_1},\ldots,v_{q_N}|(q_1,\ldots,q_N)~\text{is~every~permutation~of~} (1,\ldots,N) \}.$ $\text{E}_{\text{all}}(\cdot)$ and $\text{Var}_{\text{all}}(\cdot)$ refer to the expectation and variance under  permutations of $N!$ paths.

The rest of this paper is organized as follows. Section 2 proposes two edge-count statistics and a  relevance analysis. The power analysis is given in Section 3. Section 4 compares the proposed statistics and related statistics numerically. The applications are illustrated in Section 5. Finally, we conclude with a discussion in Section 6.

\section{Edge-count statistics}

Suppose we have random vectors $\X_t$ that are indexed by the sample $G_1,\ldots,G_k$. 
We consider the problem of testing the hypothesis: 
\begin{equation}\label{hypo}
H_0: \bm{\theta}(F_{G_1})=\cdots=\bm{\theta}(F_{G_k})\quad\text{vs}\quad H_1: \exists m\neq\ell, \bm{\theta}(F_{G_m})\neq\bm{\theta}(F_{G_\ell}).
\end{equation}

We treat vectors $\X_t$ as nodes or points $v_t$ for $t=1,\ldots,N$.
Consider a graph $\mathcal{G}$ that is a path $\mathcal{P}$ with   edges $(v_i, v_{i+1})$ for $i=1,\ldots,N-1$. The path $\mathcal{P}$ can be considered as $(v_1,\ldots,v_N)$ or $(v_N,\ldots,v_1)$.   The number of edges connecting any two sets of nodes between $G_m$ and $G_\ell$ is defined as follows:
\begin{equation}\label{Count0}
S_{\mathcal{P}}(G_m, G_\ell)=\sum_{i=1}^{N-1}I\left[\left\{(v_i\in G_m)\cap(v_{i+1}\in G_\ell)\right\} \cup\left\{(v_i\in G_\ell)\cap(v_{i+1}\in G_m)\right\}\right],
\end{equation}
 where $1\leq m,\ell\leq k$ and $I(\cdot)$ is an indicator function that takes 1 if true and 0 otherwise.

Given a path $\mathcal{P}$, for any two samples $G_m$ and $G_\ell$,   $S_{\mathcal{P}}(G_m, G_\ell)$  counts the edges between samples  for $m\neq \ell$ or within samples for $m=\ell$. We first illustrate this edge counting with an example, and then give some properties of it.

{\bf Example 1.} Suppose  there are two samples $G_1=\{1,2,3,4\}$ and $G_2=\{5,6,7,8\}$. In light of the relevance analysis, we divide each sample  into two subsamples $X_A=\{1,2\}$, $X_B=\{3,4\}$, $Y_A=\{5,6\}$, and $Y_B=\{7,8\}$. Suppose we have a path $$\mathcal{P}=(2,3,4,1,5,7,8,6).$$ 

  \begin{figure}[htb!]
    \centering
    \subfloat[\centering Rainbow representation]{{\includegraphics[width=7.5cm]{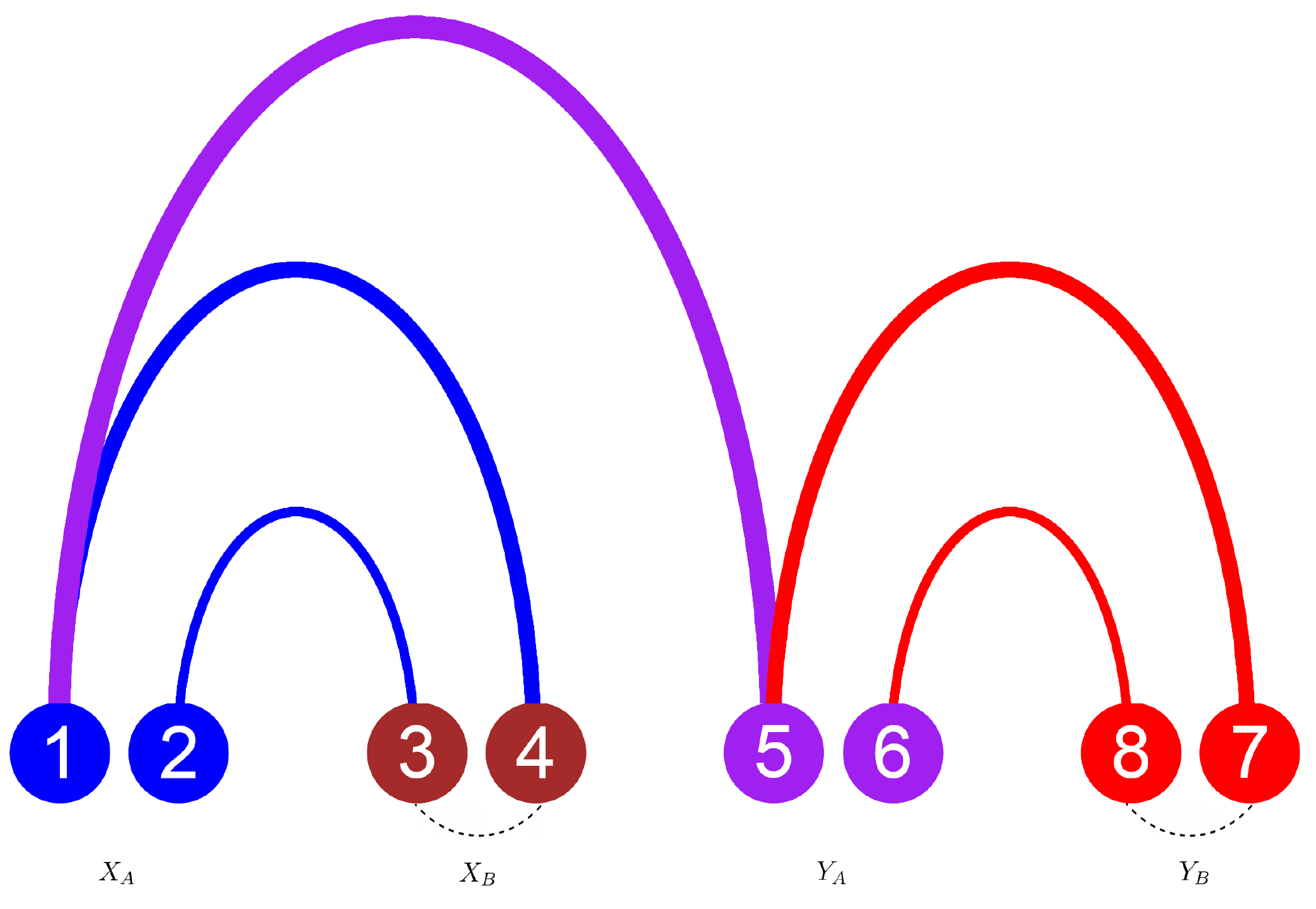} }}%
    \qquad
    \subfloat[\centering Matrix representation]{{\includegraphics[width=7.5cm]{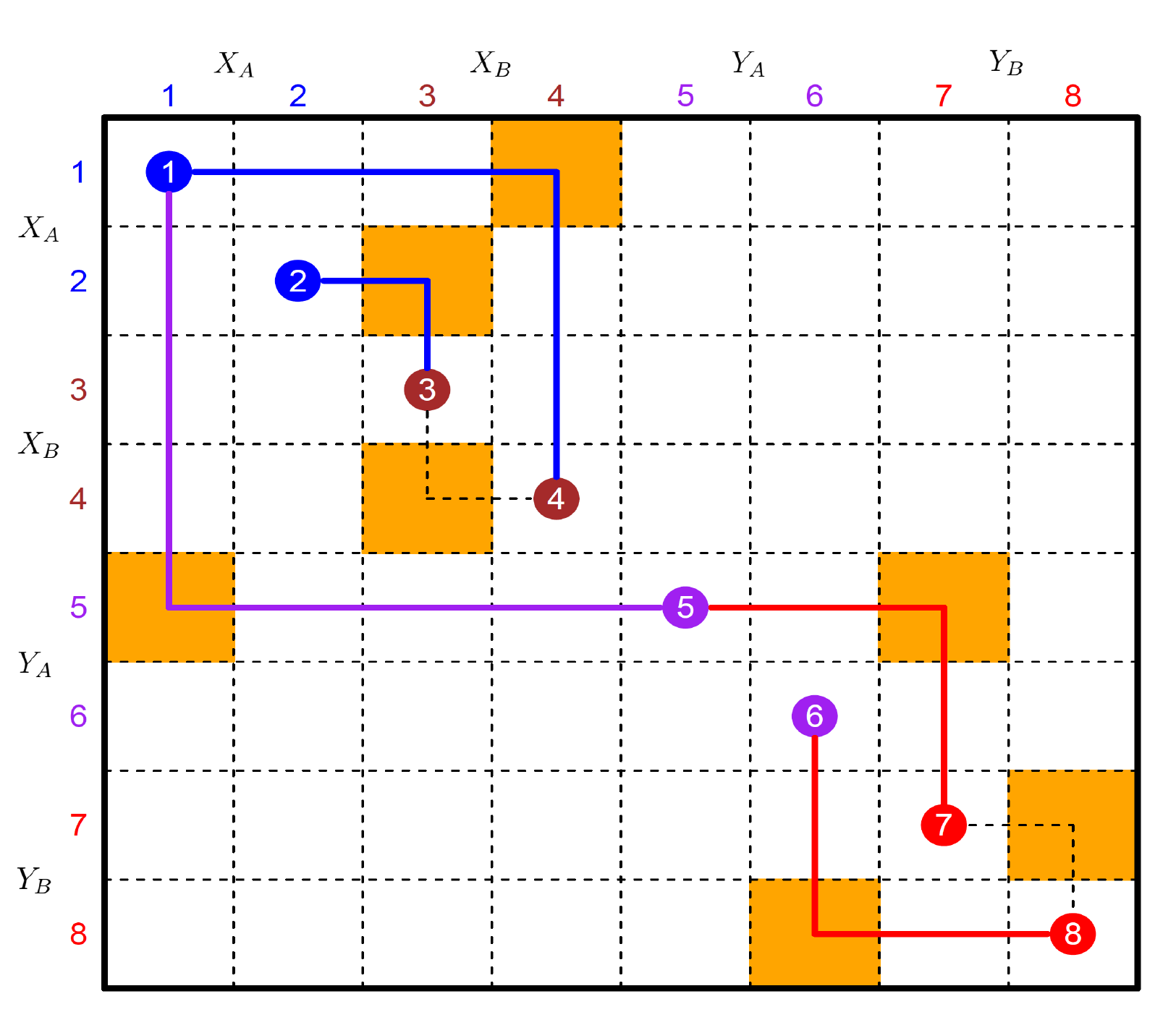} }}%
    \caption{Visualizations of edge counts}%
    \label{fig:illus}%
\end{figure}

Fig. \ref{fig:illus} provides two representations of these connected edges, namely the rainbow and matrix representations. Four subsamples $X_A,X_B,Y_A$ and $Y_B$ are   blue, brown, purple and red, respectively. Only the edges  connecting different subsamples are colored, otherwise are connected by dashed lines. It can be seen that 
$S_{\mathcal{P}}(X_A, X_B)=\#\{(1,4), (2,3)\}=2$, $S_{\mathcal{P}}(X_A, Y_A)=\#\{(1,5)\}=1$, $S_{\mathcal{P}}(X_A, Y_B)=0$, 
$S_{\mathcal{P}}(X_B, Y_A)=0$, $S_{\mathcal{P}}(X_B, Y_B)=0$, $S_{\mathcal{P}}(Y_A, Y_B)=\#\{(5,7), (6,8)\}=2$, and $S_{\mathcal{P}}(G_1, G_2)=\#\{(1,5)\}=1.$

To study the asymptotic distribution of $S_{\mathcal{P}}(G_m, G_\ell)$  shown in \eqref{Count0} under all permutations, we provide the following properties of   edge counts.

{\bf Lemma 1.}  For any non-empty and disjoint samples $\{G_i, 1\leq i\leq k\}$ ($k\geq 2$),  we have properties:
 \begin{itemize}
\item[1.] $S_{\mathcal{P}}(G_1, G_2)=S_{\mathcal{P}}(G_2, G_1)$ (symmetric);
\item[2.] $S_{\mathcal{P}}(G_1\cup G_2, G_3)=S_{\mathcal{P}}(G_1, G_3)+S_{\mathcal{P}}(G_2, G_3)$ if $k\geq3$ (additive);
\item[3.] $\text{E}_{\text{all}}\{S_{\mathcal{P}}(G_1, G_2)\}=\frac{2n_1n_2}{N}$;
\item[4.] $\text{E}_{\text{all}}\{S_{\mathcal{P}}(G_1, G_1)\}=\frac{n_1(n_1-1)}{N}$;
\item[5.] $\text{E}_{\text{all}}\{S_{\mathcal{P}}(G_1, G_2)\}^2=\frac{2n_1n_2}{N}+\frac{2n_1n_2(n_1+n_2-2)}{N(N-1)}+\frac{4n_1(n_1-1)n_2(n_2-1)}{N(N-1)}$;
\item[6.] $\text{E}_{\text{all}}\{S_{\mathcal{P}}(G_1, G_1)\}^2=\frac{n_1(n_1-1)}{N}+\frac{2n_1(n_1-1)(n_1-2)}{N(N-1)}+\frac{n_1(n_1-1)(n_1-2)(n_1-3)}{N(N-1)}$;
\item[7.] $\text{E}_{\text{all}}\{S_{\mathcal{P}}(G_1, G_2)S_{\mathcal{P}}(G_3, G_4)\}=\frac{4n_1n_2n_3n_4}{N(N-1)}$ if $k\geq4$;
\item[8.]$\text{E}_{\text{all}}\{S_{\mathcal{P}}(G_1, G_2)S_{\mathcal{P}}(G_2, G_3)\}=\frac{2n_1n_3n_2(2n_2-1)}{N(N-1)}$ if $k\geq3$;
\item[9.]$\text{E}_{\text{all}}\{S_{\mathcal{P}}(G_1, G_1)S_{\mathcal{P}}(G_2, G_2)\}=\frac{n_1(n_1-1)n_2(n_2-1)}{N(N-1)}$.
\end{itemize}
The first two properties are obvious.  
 The proofs of the other properties are given in the Appendix. In particular, if $k=2$ and $n_1+n_2=N$,  then properties 3 and 5 are consistent with the moments shown in 
\cite{WW40}[eq. 12 and eq. 13].


 \subsection{Weighted sum statistic}

  We propose a weighted sum statistic:
\begin{equation}\label{ks}
\bar{Z}_{\mathcal{P}}=\sum_{m=1}^{k-1}\sum_{\ell=m+1}^kw_{m,\ell}S_{\mathcal{P}}(G_m, G_\ell),
\end{equation}
where $w_{m,\ell}\geq 0$. In particular, if  $k=2$, then it is the classical edge-count statistic \citep{FR79}. For multiple samples with $k>2$, it is reasonable to set  $w_{m,\ell}$  to be $\{\text{Var}_{\text{all}}(S_{\mathcal{P}}(G_m, G_\ell))\}^{-1/2}$ for unbalanced sample sizes,  or zero for   subsample analysis (see Example 2).
Its asymptotic normal distribution is established in Theorem 1, a direct result of Lemma 1.

{\bf Theorem  1.} Assume $\min_\ell n_\ell\rightarrow\infty$. 
$$\frac{\bar{Z}_{\mathcal{P}}-\text{E}_{\text{all}}(\bar{Z}_{\mathcal{P}})}{\sqrt{\text{Var}_{\text{all}}(\bar{Z}_{\mathcal{P}})}}\rightarrow_dN(0, 1),$$
where  $\text{Var}_{\text{all}}(\bar{Z}_{\mathcal{P}})=\sum_{m_1=1}^{k-1}\sum_{\ell_1=m_1+1}^k\sum_{m_2=1}^{k-1}\sum_{\ell_2=m_2+1}^kw_{m_1,\ell_1}w_{m_2,\ell_2}C_{m_1,\ell_1, m_2, \ell_2}$, $C_{m_1,\ell_1, m_2,\ell_2}$ is the covariance $\text{E}_{\text{all}}\{S_\mathcal{P}(G_{m_1},G_{\ell_1})S_\mathcal{P}(G_{m_2},G_{\ell_2})\}-\text{E}_{\text{all}}\{S_\mathcal{P}(G_{m_1},G_{\ell_1})\}\text{E}_{\text{all}}\{S_\mathcal{P}(G_{m_2},G_{\ell_2})\}$, and $\text{E}(\bar{Z}_{\mathcal{P}})=\sum_{m=1}^{k-1}\sum_{\ell=m+1}^kw_{m,\ell}\text{E}_{\text{all}}\{S_{\mathcal{P}}(G_m, G_\ell)\}$.

Theorem 1 follows by  the functional limiting theory. The expectation $\text{E}_{\text{all}}(\bar{Z}_{\mathcal{P}})$ and variance $\text{Var}_{\text{all}}(\bar{Z}_{\mathcal{P}})$  can be calculated by properties 3-6 in Lemma 1.

As we know, the SHP is the path with the lowest total cost. Finding SHP is a non-deterministic polynomial (NP) problem. A heuristic algorithm (HA) is due to 
 Biswas et al. (2014).
 HA first arranges all the edges in order of increasing cost. First, the edge with the minimum cost must be selected. Then, subsequent edges are selected one by one from the remaining list of sorted edges according to the requirements of the path. If the current edge does not form a cycle with the previously selected edges and the degree of each vertex connected by the current edge or the previously selected edges is not greater than 2, then the current edge must be selected. HA terminates when $N-1$ edges are selected. The approximate SHP is formed by the selected $N-1$ edges and is represented as a set of vertices $\mathcal P^*= (v^*_1,\ldots ,v^*_{N})$.  The observed weighted sum statistic is  $\bar{Z}_{\mathcal{P}^*}.$

Given a significance level $\alpha$, we reject the null hypothesis as shown in \eqref{hypo} if $$\bar{Z}_{\mathcal{P}^*}<\text{E}_{\text{all}}(\bar{Z}_{\mathcal{P}})-z_{\alpha} \sqrt{\text{Var}_{\text{all}}(\bar{Z}_{\mathcal{P}})},$$ where $z_{\alpha}$ is the $\alpha$th quantile of standard normal distribution.

 \subsection{Minimum statistic}

We define a minimum statistic:
\begin{equation}\label{min}
\underline{Z}_{\mathcal{P}}=\min_{1\leq m<\ell\leq k}w_{m,\ell}\left\{S_{\mathcal{P}}(G_m, G_\ell)-E_{\text{all}}(S_{\mathcal{P}}(G_m, G_\ell))\right\}.
\end{equation}
Since we need to express its distribution, we define $L(i,j)=j-i+(2k-i)(i-1)/2$ for $1\leq i<j\leq k$. Then, $1\leq L(i,j)\leq k(k-1)/2$. Given any  $1\leq\ell \leq k(k-1)/2$,  only one pair $(i_\ell,j_\ell)$ exists such that $i_\ell<j_\ell$ and  $L(i_\ell,j_\ell)=\ell$. According to the functional limiting theory, we have the following asymptotic  distribution of the minimum statistic.

{\bf Theorem 2.}  For any $x$,
\begin{align}\label{lim}
P(\underline{Z}_{\mathcal{P}}\leq x)-1+
P(\bm{Z}>x\bm{\sigma})\rightarrow0,
\end{align}
where the normal random vector
$\bm{Z}\sim N(\bm{0},\bm{\Sigma})$,   the entries of covariance matrix $\bm{\Sigma}$ are $
\Sigma_{\ell_1,\ell_2}=\text{E}_{\text{all}}\{S_{\mathcal{P}}(G_{i_{\ell_1}}, G_{j_{\ell_1}})S_{\mathcal{P}}(G_{i_{\ell_2}}, G_{j_{\ell_2}})\}-\{\text{E}_{\text{all}}S_{\mathcal{P}}(G_{i_{\ell_1}}, G_{j_{\ell_1}})\}\{\text{E}_{\text{all}}S_{\mathcal{P}}(G_{i_{\ell_2}}, G_{j_{\ell_2}})\}
$ for $1\leq\ell_1,\ell_2\leq k(k-1)/2$, and the $\ell$th ($1\leq \ell\leq k(k-1)/2$) component of $\bm{\sigma}$  is equal to $w_{i_\ell, j_\ell}$ if $w_{i_\ell, j_\ell}$ is positive, otherwise the $\ell$th component is $-\infty$.

The proof of Theorem 2 is intuitive because $1-P(\underline{Z}_{\mathcal{P}}\leq x)$ can be expressed as the probability of a multivariate normal variable.
Unlike the weighted sum statistic, the minimum statistic is based on the standardized value of $S_{\mathcal{P}}(G_m, G_\ell)$.

The numerical computation of $P(\bm{Z}>x\bm{\sigma})$ is implemented by the R function \textit{pmvnorm} \citep{Ge92}. 
Assume that the critical values $\underline{z}_\alpha$ satisfies $1-P(\bm{Z}>\underline{z}_\alpha\bm{\sigma})=\alpha$.
Given a significant level $\alpha$, we reject the null hypothesis when $\underline{Z}_{\mathcal{P}^*}\leq \underline{z}_\alpha$.

\subsection{Relevance analysis}

We define the   $z$-score as
\begin{equation}\label{zs}
z_{\{m\},\{\ell\}}=\frac{S_{\mathcal{P}}(G_m, G_\ell)-\text{E}_{\text{all}}(S_{\mathcal{P}}(G_m, G_\ell))}{\sqrt{\text{Var}_{\text{all}}(S_{\mathcal{P}}(G_m, G_\ell))}},
\end{equation}
for $1\leq m\neq \ell\leq k$.  The defined $z$-score is a  nonparametric measure  of  the degree of relevance based on pairs of two samples. In particular, if $k=2$ for two samples, then both the weighted sum statistic and the minimum statistic are related to the $z$-score between the two samples. For combined samples, we can define
  \begin{equation}\label{zs2}
|z_{\mathcal{A}_1,\mathcal{A}_2}|=\frac{|S_{\mathcal{P}}(\cup_{m\in\mathcal{A}_1}G_m, \cup_{\ell\in\mathcal{A}_2}G_\ell)-\text{E}_{\text{all}}(\cup_{m\in\mathcal{A}_1}G_m, \cup_{\ell\in\mathcal{A}_2}G_\ell))|}{\sqrt{\text{Var}_{\text{all}}(\cup_{m\in\mathcal{A}_1}G_m, \cup_{\ell\in\mathcal{A}_2}G_\ell))}},
\end{equation}
 where $\mathcal{A}_1$ and $\mathcal{A}_2$ are disjoint   subsets of  $\{1,\ldots,k\}$.

 To illustrate the relevance analysis, we give the  following example.

 {\bf Example 2.}  
 The motivation for this example is 
 Simpson's paradox, where a trend appears in several different samples of data, but disappears or reverses when these samples are combined together.
Here, we consider four samples, as shown in Table  \ref{table:1}.

\begin{table}[h!]
\centering
\begin{tabular}{ |c|c|c|c|c|  }
 \hline
Samples&$G_1$&$G_2$&$G_3$&$G_4$\\\hline
Sizes&20&24&26&28\\ \hline
Distributions with $d=1000$&$N_d(\bm{0}, \mathbb{I})$&$N_d(0.01+\bm{0}, 1.1\mathbb{I})$&$N_d(0.01+\bm{0}, 1.1\mathbb{I})$&$N_d(\bm{0}, \mathbb{I})$\\
 \hline
\end{tabular}
\caption{Individuals are distributed along each sample.}
\label{table:1}
\end{table}

\begin{figure}[htb!]
\begin{center}
\includegraphics[width=\textwidth]{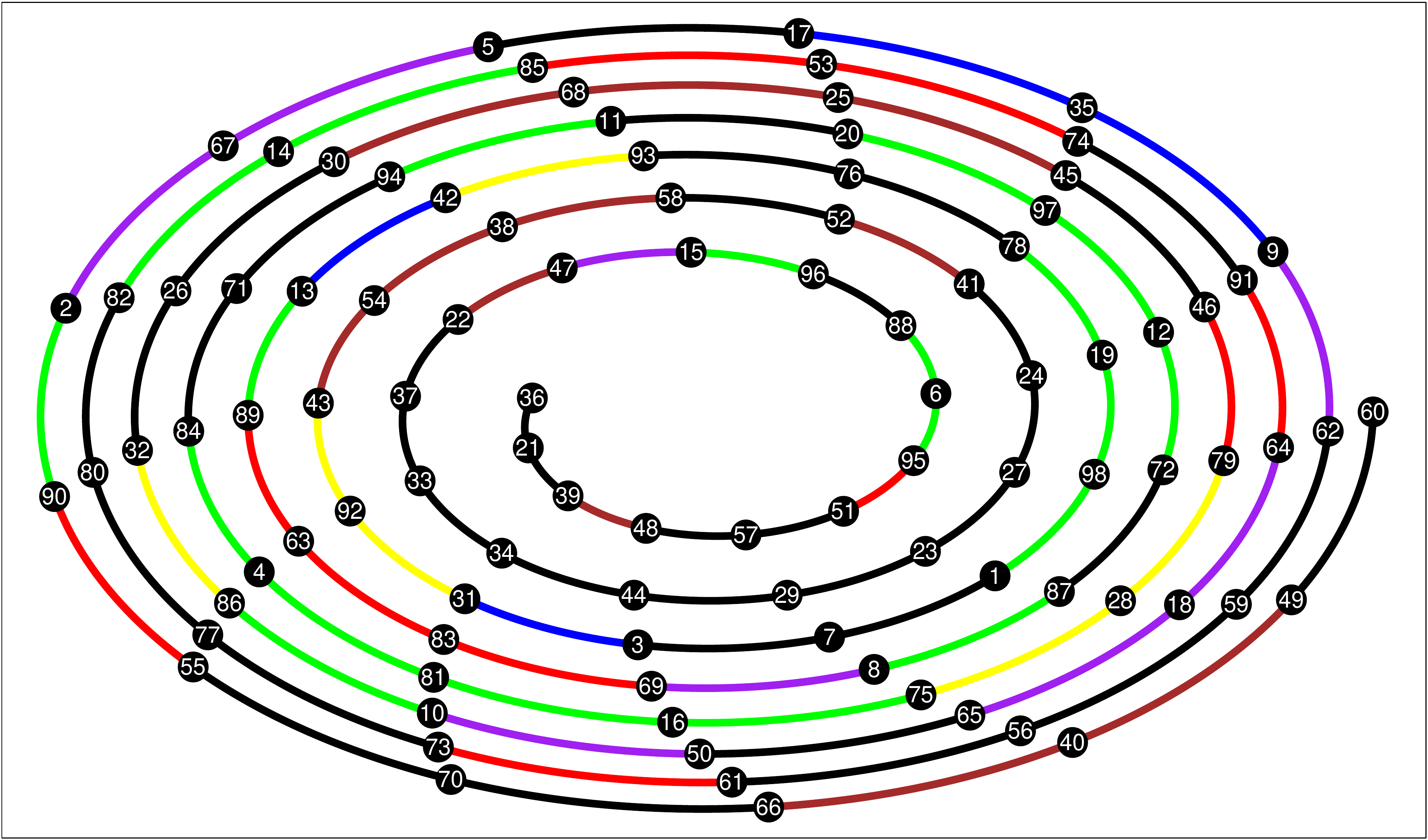}
\caption{Fifty nine colored edges between samples and thirty nine black edges within samples.}
\label{fig.illus}
\end{center}
\end{figure}

We draw data along each sample according to Table \ref{table:1}.
The approximate SHP, $\mathcal P^*$, is plotted  in Figure \ref{fig.illus}, where four blue edges connect samples $G_1$ and $G_2$, eight purple edges connect samples $G_1$ and $G_3$, twenty green edges connect samples $G_1$ and $G_4$, eleven brown edges connect samples $G_2$ and $G_3$, six yellow edges connect samples $G_2$ and $G_4$, and ten red edges connect samples $G_3$ and $G_4$. The $z$-scores of these edge counts are listed in  Table \ref{table:2}.

\begin{table}[h!]
\centering
\begin{tabular}{ |c|c|c|c| }
 \hline
&$G_2$&$G_3$&$G_4$\\\hline
$G_1$&-2.3&-1.0&3.2\\\hline
$G_2$&&-0.6&-2.6\\\hline
$G_3$&&&-1.6\\\hline 
\end{tabular}
\caption{The $z$-scores of edge counts for measuring the degree of relevance.}
\label{table:2}
\end{table}

From  Table \ref{table:2}, the four smallest $z$-scores are: $z_{\{2\},\{4\}}=-2.6$,
$z_{\{1\},\{2\}}=-2.3$,
$z_{\{3\},\{4\}}=-1.6$, $z_{\{1\},\{3\}}=-1.0$. They   correspond to different samples. Suppose we want to compare the difference between combined samples: $G_1\cup G_2$ and $G_3\cup G_4$, ignoring the within-sample differences, i.e., test 
$$H_0: \bm{\theta}(F_{G_1\cup G_2})=\bm{\theta}(F_{G_3\cup G_4})\quad \text{vs}\quad H_1: \bm{\theta}(F_{G_1\cup G_2})\neq\bm{\theta}(F_{G_3\cup G_4}).$$
The combined $z$-score $z_{\{1,2\},\{3,4\}}$ is -0.7. In addition, both the weighted sum test \eqref{ks} and  the minimum test \eqref{min} give the same $p$-values of 0.237.  To make   meaningful comparisons, we test for the difference  between $G_1$ and $G_3$ or  between $G_2$ and $G_4$, i.e.,
 $$H_0: \bm{\theta}(F_{G_1})=\cdots=\bm{\theta}(F_{G_4})\quad \text{vs}\quad H_1: \bm{\theta}(F_{G_1})\neq\bm{\theta}(F_{G_3})\quad\text{or}\quad \bm{\theta}(F_{G_2})\neq\bm{\theta}(F_{G_4}).$$
We  set the weights $w_{1,2}$, $w_{1,4}$, $w_{2,3}$ and $w_{3,4}$ to be zeros  and   focus only on the purple and yellow edges  in Figure \ref{fig.illus}. The weighted sum test gives a $p$-value of 0.009, while the  minimum test give a  $p$-value of 0.007. This comparison is more meaningful. Thus,   differences between samples are apparent but disappear when these samples are combined.

\section{Power}

In this section, we  discuss the effect of edge costs on   power. 
Denote  the cost between $\X_{t_1}$ and $\X_{t_2}$   as $\D(t_1, t_2)$ and its limit  as $\D^*(t_1, t_2)$. We make the following regular assumptions.

 \begin{itemize}
    \item[A1.] $\D(t_1, t_2)>0, t_1\neq t_2,~\text{almost~ surely.}$
    \item[A2.] $\D(t_1, t_2)=\D(t_2, t_1) ~(\text{symmetry}).$
    \item[A3.] $\D(t_1, t_2)\leq\D(t_1, t_3)+\D(t_2, t_3) ~(\text{triangular~ inequality}).$
    \item[A4.] $\D(t_1, t_2)-\D^*(g_{t_1}, g_{t_2})=\O_p(d^{-1/2})~\text{and}~Nd^{-1/2}\rightarrow0$ (rates).
    \item[A5.] If $\bm{\theta}(F_{g_{t_1}})\neq \bm{\theta}(F_{g_{t_2}}), ~\text{then}~\D^*(g_{t_1}, g_{t_2})>\min\{\D^*(g_{t_1}, g_{t_1}),\D^*(g_{t_2}, g_{t_2})\}$ (increased cost).
\end{itemize}

Here, we consider continuous random vectors and therefore require A1. 
Because we actually apply the permutation test, we  need the symmetric condition A2. 
  Due to the ordering of costs, we need A3.
Since we consider that $d$ converges to $\infty$, by the law of large numbers it is natural to assume that the limit of cost exists: $\D(t_1, t_2)\rightarrow_p\D^*(g_{t_1}, g_{t_2})$.
Since there are  $N(N-1)/2$ random  costs, we further assume some rates in A4.
   These reflect  the relationship between the sample size $N$ and dimension $d$  and  lead  to
$$P\left\{\cup_{1\leq t_1<t_2\leq N}|\D(t_1, t_2)-\D^*(g_{t_1}, g_{t_2})|>\varepsilon\right\}=\O(N^2d^{-1})\rightarrow0,$$ for any $\varepsilon>0$.
To obtain power, we require the between-sample cost must be greater than one of within-sample costs in A5.

Usually we assign the $\gamma$-norm as the cost:  for $0<\gamma\leq2$,
 \begin{equation}\label{GN}
 \D(t_1, t_2)=d^{-1/\gamma}||\X_{t_1}-\X_{t_2}||_\gamma.
 \end{equation}

When $\gamma=2$,
the $2$-norm is   the scaled Euclidean distance, $\D(t_1, t_2)=d^{-1/2}||\X_{t_1}-\X_{t_2}||_2$.
It is obvious that A1-A3 are satisfied. A4 can be confirmed by using the Markov inequality under some weak conditions. In particular, if the components of the random vectors $\X_{t}$  are 
independent and identically distributed (IID),  and $\bm{\theta}(F_{g_t})$ is a parametric vector including mean and variance, denoted as $(\mu_{g_t}, \sigma^2_{g_t})^\top$, then   $\D^*(g_{t_1}, g_{t_1})=\sqrt{2\sigma^2_{g_{t_1}}}$,   $\D^*(g_{t_2}, g_{t_2})=\sqrt{2\sigma_{g_{t_2}}}$, and $\D^*(g_{t_1}, g_{t_2})=\sqrt{(\mu_{g_{t_1}}-\mu_{g_{t_2}})^2+\sigma^2_{g_{t_1}}+\sigma^2_{g_{t_2}}}$. As $\sqrt{\sigma^2_{g_{t_1}}+\sigma^2_{g_{t_2}}}\geq \min\{\sqrt{2\sigma^2_{g_{t_1}}},\sqrt{2\sigma^2_{g_{t_2}}}\}$, $\D^*(g_{t_1}, g_{t_2})>\min\{\D^*(g_{t_1}, g_{t_1}),\D^*(g_{t_2}, g_{t_2})\}$ when $\mu_{g_{t_1}}\neq\mu_{g_{t_2}}$ or $\sigma^2_{g_{t_1}}\neq\sigma^2_{g_{t_2}}$. This implies that A5 holds.

Next, we consider other values of $\gamma$:  $\D(t_1, t_2)=d^{-1/\gamma}||\X_{t_1}-\X_{t_2}||_\gamma$ for $0<\gamma<2$. In particular, if $\gamma=1$, that is the scaled Mahalanobis distance that is widely used in cluster analysis and classification. We proceed to the case of IID.  
Similarly, we can verify  the conditions A1-A4. 
Since  $\D^*$ has no closed form,  we apply the results of \cite{SR05}   to prove A5, which is shown in Lemma 2.

{\bf Lemma 2}. 
Suppose  the components of the random vectors $\X_{t}$  are  IID. If $F_{g_{t_1}}\neq F_{g_{t_2}}$, then $2\{\D^*(g_{t_1}, g_{t_2})\}^\gamma>\{\D^*(g_{t_1}, g_{t_1})\}^\gamma+\{\D^*(g_{t_2}, g_{t_2})\}^\gamma$.
 
By Lemma 2, $\{\D^*(g_{t_1}, g_{t_2})\}^\gamma>\min\{\{\D^*(g_{t_1}, g_{t_1})\}^\gamma,\{\D^*(g_{t_2}, g_{t_2})\}^\gamma\}$   that leads to A5.

If we only need to detect  changes in the common mean, we can consider the average cost 
\begin{equation}\label{AW}
\overline\D(t_1, t_2)=d^{-1}||X_{t_1\bullet}-X_{t_2\bullet}||_1,
\end{equation} 
which was applied in Shi, Wu and Rao (2018). It is clear that $\overline\D(t_1, t_2)$ satisfies A1-A5.

When the covariances may not be equal, we can consider another cost to capture the change in mean or covariance:
\begin{equation}\label{CN}
\dot{\D}(t_1, t_2)=d^{-1/2}\sqrt{||\X_{t_1}-\X_{t_2}||^2_2+||\dot{\X}_{t_1}||^2_2+||\dot{\X}_{t_2}||^2_2}.
\end{equation}
This new cost obviously  satisfies A1-A4. 
To verify the condition   A5, we consider two autoregressive (AR) processes $X_{{{t_\ell}},j}=\mu_{{t_\ell}}+\phi_{g_{t_\ell}} X_{{{t_\ell}},j-1}+\varepsilon_j$, where $|\phi_{g_{t_\ell}}|<1$, $\varepsilon_j$ is white noise with variance $\sigma^2_{g_{t_\ell}}$, and $\ell=1,2$. We have that $||\X_{t_1}-\X_{t_2}||^2_2$ converges to $$(\mu_{g_{t_1}}-\mu_{g_{t_2}})^2+\frac{\sigma_{g_{t_1}}^2}{1-\phi_{g_{t_1}}^2}+\frac{\sigma_{g_{t_2}}^2}{1-\phi_{g_{t_2}}^2}.$$
 Moreover, it can be seen that $X_{{t_\ell},j}-X_{{t_\ell},j-1}$ are  autoregressive moving average (ARMA) processes with mean AR coefficient $\phi_\ell$ and MA coefficient $-1$. Thus, $||\dot{\X}_{{t_\ell}}||^2_2$ converges to
  $$\frac{(2-2\phi_{g_{t_\ell}})\sigma^2_{g_{t_\ell}}}{1-\phi_{g_{t_\ell}}^2}.$$
  
 Now, we write  $\dot{\D}^*(g_{t_1}, g_{t_2})=\sqrt{(\mu_{g_{t_1}}-\mu_{g_{t_2}})^2+\frac{(3-2\phi_{g_{t_1}})\sigma^2_{g_{t_1}}}{1-\phi_{g_{t_1}}^2}+\frac{(3-2\phi_{g_{t_2}})\sigma^2_{g_{t_2}}}{1-\phi_{g_{t_2}}^2}}$. Furthermore, we write 
  $\dot{\D}^*(g_{t_1}, g_{t_1})=\sqrt{\frac{(6-4\phi_{g_{t_1}})\sigma^2_{g_{t_1}}}{1-\phi_{g_{t_1}}^2}}$, and $\dot{\D}^*(g_{t_2}, g_{t_2})=\sqrt{\frac{(6-4\phi_{g_{t_2}})\sigma^2_{g_{t_2}}}{1-\phi_{g_{t_2}}^2}}$.   
   When $\mu_{g_{t_1}}\neq\mu_{g_{t_2}}$ or  
   $\frac{(3-2\phi_{g_{t_1}})\sigma^2_{g_{t_1}}}{1-\phi_{g_{t_1}}^2}\neq\frac{(3-2\phi_{g_{t_2}})\sigma^2_{g_{t_2}}}{1-\phi_{g_{t_2}}^2}$,  
  $\dot{\D}^*(g_{t_1}, g_{t_2})>\min\{\dot{\D}^*(g_{t_1}, g_{t_1}), \dot{\D}^*(g_{t_2}, g_{t_2})\}$.

%
%
 
 To show how   power is obtained, we first provide an upper bound of the count $S_{\mathcal{P}}(g_{t_1}, g_{t_2})$.
 
{\bf Theorem 3.} Assume that  conditions  A1-A5 are satisfied.  If   $\bm{\theta}(F_{g_{t_1}})\neq \bm{\theta}(F_{g_{t_2}})$, then $S_{\mathcal{P}^*}(g_{t_1}, g_{t_2})\leq 2$ in probability as $d\rightarrow\infty$.

The proof is placed in the Appendix. The upper bound is determined by the degree constraint of the vertices. We remark that one can equivalently consider a
 degree-constrained minimum spanning tree
(DCMST) where the maximum vertex degree is limited to a
certain constant 2.

 {\bf Theorem 4.} Assume that  conditions  A1-A5 are satisfied. If   $\lim_{N\rightarrow\infty}n_i/N>0$  and there exists at least one pair $(i, j)$ such that $w_{i,j}>0$ and $\bm{\theta}(F_{G_i})\neq \bm{\theta}(F_{G_j})$.    The power of the $k$-sample test based on $\bar{Z}_{\mathcal{P}}$ in \eqref{ks} or $\underline{Z}_{\mathcal{P}}$ in \eqref{min} tends to 1 as $d\rightarrow\infty$.

The proof is placed in the Appendix. Theorem 4 justifies why the proposed $k$-sample tests can obtain power.



%


\section{Simulations}

\subsection{Two-sample comparison}
We generate data   $\{\bm{X}_t, 1\leq t\leq 60\}$ from $N_d(\bm{\mu}_t,\bm{\Sigma}_t)$. 
We consider two samples $G_1$ and $G_2$ whose sizes are $n_1=20$ and $n_2=40$, respectively. We set the parameters:
\begin{equation*}
\bm{\mu}_t=\begin{cases}\bm{\mu}^{(1)}, 1\leq t\leq n_1,\\
\bm{\mu}^{(2)}, n_1<t\leq n_1+n_2,
\end{cases}\quad \bm{\Sigma}_t=\begin{cases}\bm{\Sigma}^{(1)}, 1\leq t\leq n_1,\\
\bm{\Sigma}^{(2)}, n_1<t\leq n_1+n_2.
\end{cases}
\end{equation*}

To examine the effect of parameters on   power, we consider the following three cases.

Case 1. $\bm{\mu}^{(1)}=\bm{\mu}^{(2)}-0.1=\bm{0}$ and $\bm{\Sigma}^{(1)}_{i,j}=\bm{\Sigma}^{(2)}_{i,j}=0.2^{|i-j|}$.
 
Case 2. $\bm{\mu}^{(1)}=\bm{\mu}^{(2)}=\bm{0}$, $\bm{\Sigma}^{(1)}_{i,j}=0.2^{|i-j|}$, and $\bm{\Sigma}^{(2)}_{i,j}=0.4^{|i-j|}$.

Case 3. $\bm{\mu}^{(1)}=\bm{\mu}^{(2)}-0.1=\bm{0}$,   $\bm{\Sigma}^{(1)}_{i,j}=0.2^{|i-j|}$, and $\bm{\Sigma}^{(2)}_{i,j}=0.4^{|i-j|}$.

The original test was proposed by \cite{BS96}   and denoted as T-1. Some variants were proposed by  \cite{CQ10}    and \cite{CLX14}  denoted as T-2 and T-3, respectively. These variants can   treat equal covariances for case 1 and unequal convariances for cases 2 and 3, respectively. Four MST-based tests have been proposed by \cite{FR79}, \cite{CF17}, \cite{CCS18}, \cite{ZC21}, which are denoted by T-4, T-5, T-6, and T-7, respectively. The latest test to be compared was proposed by \cite{MW20}, which is denoted as K-1.

Since we are considering here a comparison of two samples, both
weighted sum statistic \eqref{ks} and minimum statistic \eqref{min}  are equivalent. We only consider 
the weighted sum statistic \eqref{ks}.
We consider the cost   \eqref{GN} for   $\gamma=2$, denoted by K-2. For a fair comparison, we apply the cost \eqref{AW} in case 1 and the cost \eqref{CN} in cases 2 and 3, which we denote both as K3.

\begin{figure}[htb!]
\begin{center}
\includegraphics[width=\textwidth]{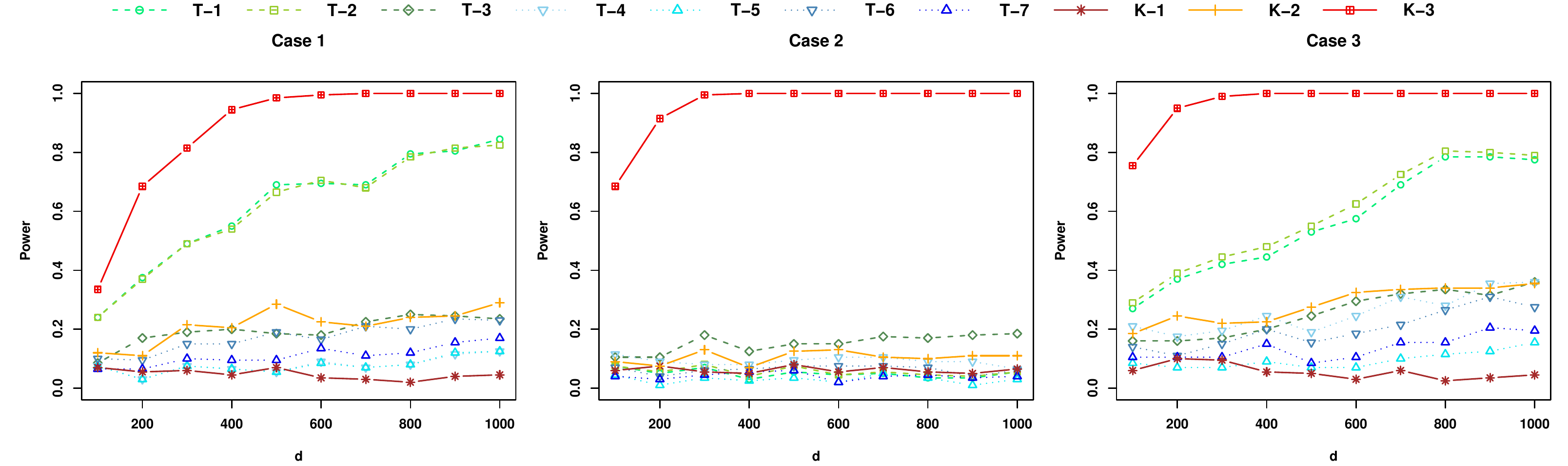}
\caption{Power comparison of two-sample tests.}
\label{fig.power2}
\end{center}
\end{figure}

Fig. \ref{fig.power2} shows the estimated power,  a percentage in the 200 trials (\%) when the null hypothesis is
rejected at the 0.05 level for each of the two-sample tests. We can see that
  K-3 has highest power; T-1 and T-3 are comparable and can gain power when the mean value changes; the others have little powerful; and overall, K-2 does not perform poorly.


%


\subsection{Three-sample comparison} 

Similar to the two-sample setup, we allow $t$ to vary from 1 to 90.  We  consider three samples $G_1$, $G_2$ and $G_3$ with sizes $n_1=20$, $n_2=30$  and $n_3=40$, respectively. We set the parameters:

\begin{equation*}
\bm{\mu}_t=\begin{cases}\bm{\mu}^{(1)}, 1\leq t\leq n_1,\\
\bm{\mu}^{(2)}, n_1<t\leq n_1+n_2,\\
\bm{\mu}^{(3)}, n_1+n_2<t\leq n_1+n_2+n_3,
\end{cases}\quad \bm{\Sigma}_t=\begin{cases}\bm{\Sigma}^{(1)}, 1\leq t\leq n_1,\\
\bm{\Sigma}^{(2)}, n_1<t\leq n_1+n_2,\\
\bm{\Sigma}^{(3)}, n_1+n_2<t\leq n_1+n_2+n_3.\\
\end{cases}
\end{equation*}

We consider another three cases.

 Case 4. $\bm{\mu}^{(1)}=\bm{\mu}^{(2)}=\bm{\mu}^{(3)}-0.1=\bm{0}$, $\bm{\Sigma}^{(1)}_{i,j}=\bm{\Sigma}^{(2)}_{i,j}=0.2^{|i-j|}$, and $\bm{\Sigma}^{(3)}_{i,j}=0.4^{|i-j|}$.
 
Case 5. $\bm{\mu}^{(1)}=\bm{\mu}^{(2)}=\bm{\mu}^{(3)}-0.1=\bm{0}$, $\bm{\Sigma}^{(1)}_{i,j}=0.2^{|i-j|}$, $\bm{\Sigma}^{(2)}_{i,j}=0.4^{|i-j|}$, and $\bm{\Sigma}^{(3)}_{i,j}=0.6^{|i-j|}$.

Case 6. $\bm{\mu}^{(1)}=\bm{\mu}^{(2)}+0.1=\bm{\mu}^{(3)}-0.1=\bm{0}$, $\bm{\Sigma}^{(1)}_{i,j}=0.2^{|i-j|}$, $\bm{\Sigma}^{(2)}_{i,j}=0.4^{|i-j|}$, and $\bm{\Sigma}^{(3)}_{i,j}=0.6^{|i-j|}$.

Since K-1 can be applied for multiple-sample comparison, we include it here for further comparison. 
Here,  the
weighted sum statistic \eqref{ks} and the minimum statistic \eqref{min}  are not the same. For the weighted sum statistic \eqref{ks}, we 
 consider the cost   \eqref{GN} with  $\gamma=2$ denoted by K-2 and the cost   \eqref{CN} denoted as K3, while for minimum statistic \eqref{min}, we 
 consider the cost   \eqref{GN} with  $\gamma=2$ denoted by K-4 and the cost   \eqref{CN} denoted as K5.

\begin{figure}[htb!]
\begin{center}
\includegraphics[width=\textwidth]{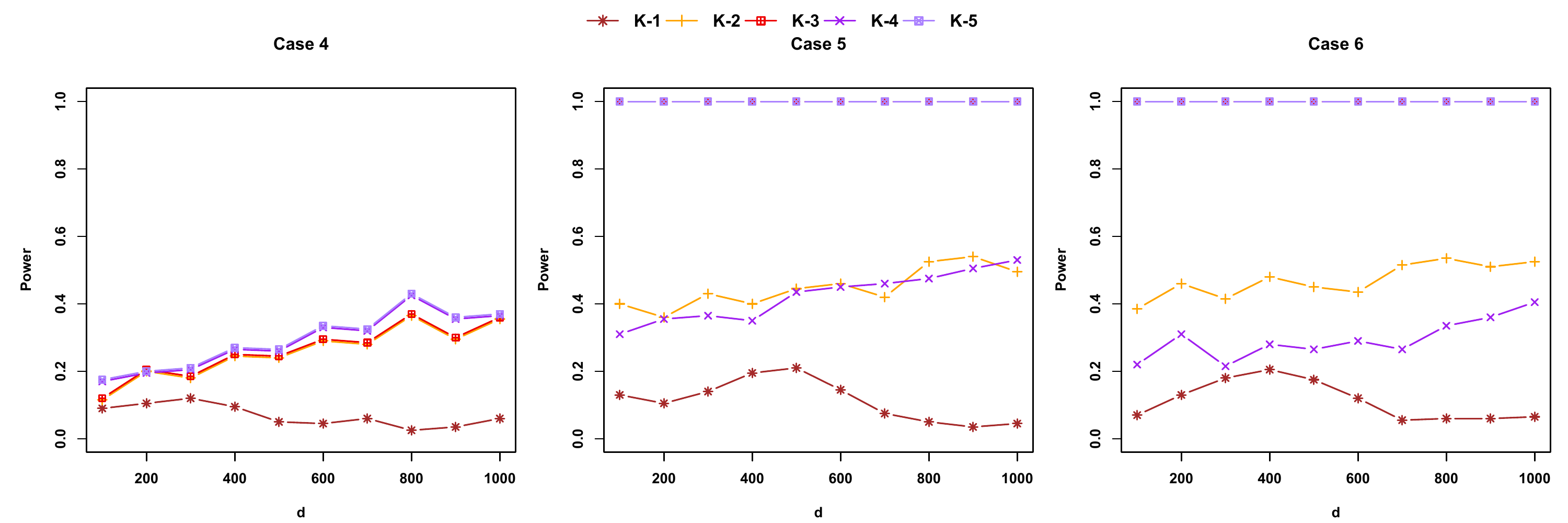}
\caption{Power comparison of three-sample tests.}
\label{fig.power3}
\end{center}
\end{figure}

Similarly, Fig. \ref{fig.power3} presents the estimated power. It can be seen that 
  all of our tests are comparative in case 4; K-3 and K-5 obtain more power in cases 5 and 6;   and  K-1 has a worse performance.

\section{Real data analysis}

\subsection{Microarray data analysis}  

We analyze the tumor data $\{\bm{X}_t\in R^d, 1\leq t\leq 801, d=20,531\}$ indexed to five samples of BRCA, COAD, KIRC, LUAD and PRAD. The sizes of all samples are 300, 78, 146, 141, and 136.   Using the proposed relevance analysis, we compared gene expression levels across 5 tumor types.  
The following table lists the  $z$-scores  between each two samples, as in \eqref{zs}.

\begin{table}[ht]
\def~{\hphantom{0}}
\caption{The $z$-scores  between each two samples, as in \eqref{zs}.}
 \centering
\begin{tabular}{|l|l|l|l|l|}\hline
     & COAD& KIRC& LUAD& PRAD\\\hline
 BRCA& -9.86&-13.62&-12.72&-12.84\\\hline
 COAD&&-6.13& -5.13& -5.88 \\\hline
 KIRC&&&-8.54&  -8.36\\\hline
 LUAD&&&&-8.19\\\hline
\end{tabular}
\label{table1}
\end{table}

From   Table \ref{table1}, we can see that the two types of tumors BRCA and KIRC are mostly irrelevant, and the other two types of tumors COAD and LUAD are mostly relevant. Both the weighted sum statistic \eqref{ks} and minimum statistic \eqref{min} return extremely small $p$-values.


To compare the relevance for  significant components, we perform a one-way ANOVA test on each component. The $p$-value of each test is adjusted by the Benjamini-Hochberg correction \citep{BH95}.
We select 19,565 significant components whose tests have $p$-values less than 0.05. Table \ref{table1.2} presents the relevance analysis based on the  significant components.

\begin{table}[ht]
\def~{\hphantom{0}}
\caption{The $z$-scores  between each two samples on  significant components, as in \eqref{zs}.}
 \centering
\begin{tabular}{|l|l|l|l|l|}\hline
     & COAD& KIRC& LUAD& PRAD\\\hline
 BRCA& -9.69&-13.74&-12.97&-13.10\\\hline
 COAD&&-6.13& -5.35& -5.88 \\\hline
 KIRC&&&-8.37&  -8.19\\\hline
 LUAD&&&&-8.19\\\hline
\end{tabular}
\label{table1.2}
\end{table}


From   Table \ref{table1.2}, we can see the small differences compared to the   Table \ref{table1}.

\subsection{Functional magnetic resonance imaging  data  analysis}

 BOLD was measured from each subject in 7 mice, under the conditions of   9.4 T and 15.2 T MR systems. The output of this measurement wass a 3D volume of each subject, where each voxel contained   pixel values with dimensions of 228$\times$160$\times132$. All of these were concatenated into a 4D image, where the fourth dimension was the time point, $t$, varying from 1 to 120. The time period includes 40-s pre-stimulus, 20-s stimulus, and 60-s post-stimulus.

\begin{figure}[htb!]
\begin{center}
\includegraphics[width=\textwidth]{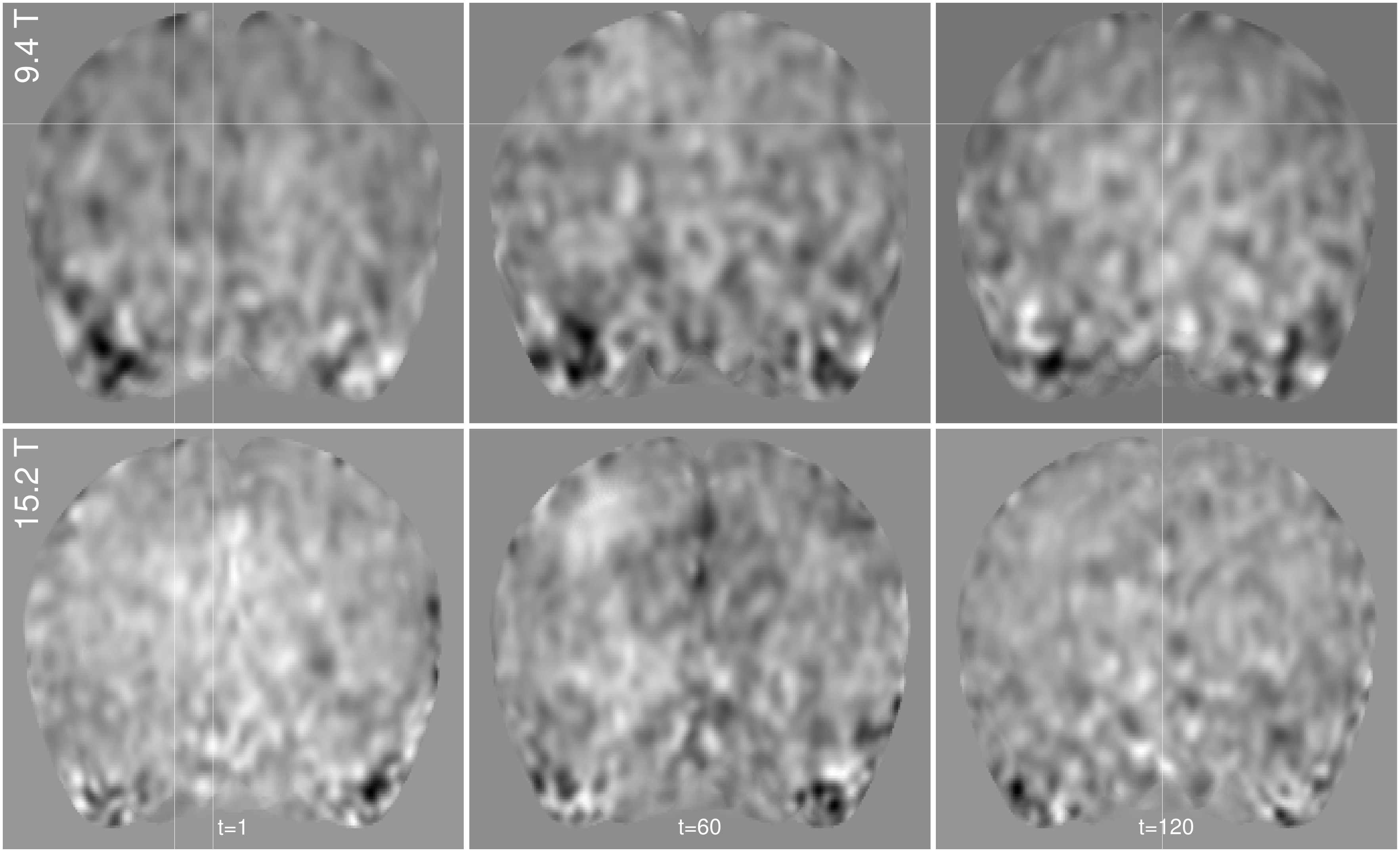}
\caption{Multi-slice averaged fMRI maps of one animal projected along one of the dimensions.}
\label{fig.brain}
\end{center}
\end{figure}

To obtain the differences in conditions among all subjects,
we first average the 3D voxels  along the third dimension.  Fig. \ref{fig.brain} presents the projected 2D voxels for two conditions (9.4 T and 15.2 T) and three time points ($t$=1, 60, 120).
We then take the differences in the projected 2D voxels for each subject across  conditions, and finally consider the cummulative differences for all 7 subjects.  
We   convert each   cumulative difference matrix into 
a vector of length 228$\times$160, which consists of all columns of the original matrix. Let $\bm{X}_t$ represent  the vector of time point $t$.

Table \ref{table2} provides the   $z$-score values in the  pre-stimulus, stimulus, and post-stimulus samples. Pre-stimulus and stimulus are mostly irrelevant, while stimulus and post-stimulus are mostly relevant. This can be explained by the fact that the response   increases rapidly in the stimulus period and decreases slowly   in post-stimulus period.

\begin{table}[ht]
\def~{\hphantom{0}}
\caption{The $z$-scores between each two time periods, as in \eqref{zs}.}
 \centering
\begin{tabular}{|l|l|l|}\hline
      & Stimulus& Post-stimulus\\\hline
Pre-stimulus& 3.63&2.92\\\hline
Stimulus&&0.64\\\hline 
\end{tabular}
\label{table2}
\end{table}
 The  minimum statistic-based test \eqref{min} captures the change between pre-stimulus and stimulus and returns a $p$-value of 0.0004, while the weighted sum   statistic-based test \eqref{ks}  has a $p$-value of  0.1472. One can set the weight  between
pre-stimulus and post-stimulus to be zero, ignoring the comparison between them. Then, the $p$-values based on the weighted sum statistic \eqref{ks} and the minimum statistic \eqref{min} are 0.00001 and 0.0003, respectively. Overall, these tests confirm the change in the difference between conditions from pre-stimulus to post-stimulus.

\section{Discussion}

Based on the graph method, we provide a technology to compare k high-dimensional samples. The proposed $z$-scores can be used to measure the degree of relevance  between the combined samples. We  propose two nonparametric tests and show that  powers can be obtained under weak conditions. Two datasets are analyzed for illustration.

We note that there is a connection between the k-sample problem and the multiple change point problem. Sliding window analysis can bridge them  \citep{Le19}.
We will further investigate this connection in our future work.
 
 \section{Acknowledgments}

We thank Dr. Won Beom Jung for sharing fMRI data.  The research is partially supported by the Natural Sciences and Engineering Research Council of Canada.

\section{Appendix}

\subsection{Proof of Lemma 1} 
\begin{figure}[htb!]
\begin{center}
\includegraphics[width=\textwidth]{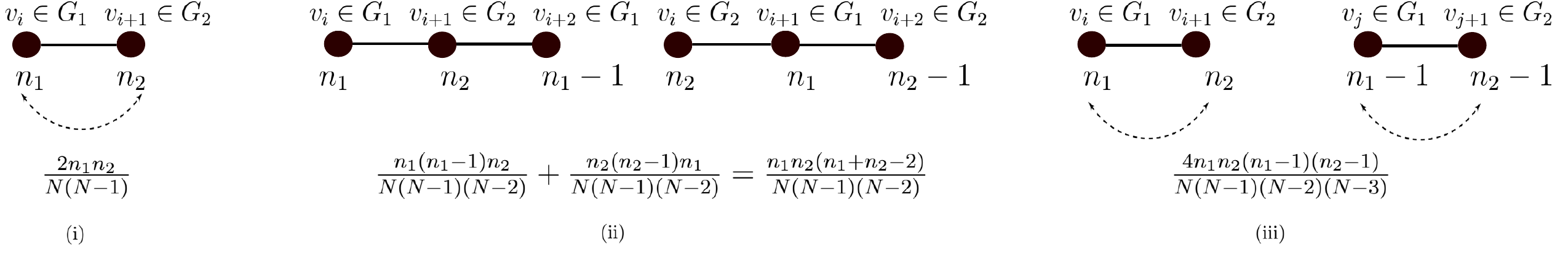}
\caption{Illustrations of calculation of probabilities.}
\label{fig.prob}
\end{center}
\end{figure}

%
%
%
%
For the first moment, we calculate
  \begin{align}\label{m1}
\text{E}_{\text{all}}\{S_{\mathcal{P}}(G_1, G_2)\}&=\sum_{i=1}^{n-1}P\left[\left\{(v_i\in G_1)\cap(v_{i+1}\in G_2)\right\} \cup\left\{(v_i\in G_2)\cap(v_{i+1}\in G_1)\right\}\right],\nonumber\\
 &=\sum_{i=1}^{n-1}\frac{2n_1n_2}{N(N-1)}=\frac{2n_1n_2}{N},
\end{align} 
where the calculation of the probability is illustrated in  
Fig. \ref{fig.prob} (i).

%
%
%
%
%

When $G_2=G_1$,
Property 4 follows by the fact that $n_2=n_1-1$. 

For the second moment, we calculate
  \begin{equation}\label{m2}
 \text{E}_{\text{all}}\{S_{\mathcal{P}}(G_1, G_2)\}^2=\sum_{i=1}^{n-1}\sum_{j=1}^{n-1}p_{i,j},
 \end{equation}
 where $p_{i,j}$ is the expectation of two indicator functions as in \eqref{Count0} for indexes $i$ and $j$, respectively.

 As shown in Fig. \ref{fig.prob}, there are three possible values of $p_{i,j}$, corresponding to three cases (i), (ii), and (iii). The first case (i) is    $\{i=j, 1\leq i,j\leq N-1\}$, which has  size   $N-1$. The second case (ii) is   $\{|i-j|=1, 1\leq i,j\leq N-1\}$, which has size $2(N-2)$. The third case (iii) is for $\{|i-j|>1, 1\leq i,j\leq N-1\}$, whose size is $(N-2)(N-3)$. Therefore,
\begin{align*}
 \text{E}_{\text{all}}\{S_{\mathcal{P}}(G_1, G_2)\}^2&=(N-1)\frac{2n_1n_2}{N(N-1)}+2(N-2)\frac{n_1n_2(n_1+n_2-2)}{N(N-1)(N-2)}\\
 &+(N-2)(N-3)\frac{4n_1n_2(n_1-1)(n_2-1)}{N(N-1)(N-2)(N-3)}.
\end{align*} 
 This is the same as the simplified  Property 5.

 Property 6 follows for the same samples $G_1=G_2$, where 
 \begin{align*}
 \text{E}_{\text{all}}\{S_{\mathcal{P}}(G_1, G_1)\}^2&=(N-1)\frac{n_1(n_1-1)}{N(N-1)}+2(N-2)\frac{n_1(n_1-1)(n_1-2)}{N(N-1)(N-2)}\\
 &+(N-2)(N-3)\frac{n_1(n_1-1)(n_1-2)(n_1-3)}{N(N-1)(N-2)(N-3)}.
\end{align*}
 
To prove Property 7, we note that   $G_1$, $G_2$, $G_3$ and $G_4$ are disjoint groups. Therefore, the values of $p_{i,j}$ for cases (i) and (ii) are zeroes. We only need to consider   case (iii). This can be done by 
  \begin{align*}
\text{E}_{\text{all}}\{S_{\mathcal{P}}(G_1, G_2)S_{\mathcal{P}}(G_3, G_4)\}&=(N-1)\times 0+2(N-2)\times0\\
 &+(N-2)(N-3)\frac{4n_1n_2n_3n_4}{N(N-1)(N-2)(N-3)}.
\end{align*}

To prove Property 8, we note that  
 $$S_{\mathcal{P}}(G_1, G_2)+S_{\mathcal{P}}(G_2, G_3)=S_{\mathcal{P}}(G_1\cup G_3, G_2).$$

On both sides of this equation, we take their squares and calculate their expected values. 

$$2\text{E}_{\text{all}}\{S_{\mathcal{P}}(G_1, G_2)S_{\mathcal{P}}(G_2, G_3)\}=E\{S_{\mathcal{P}}(G_1\cup G_3, G_2)\}^2-E\{S_{\mathcal{P}}(G_1, G_2)\}^2-E\{S_{\mathcal{P}}(G_2, G_3)\}^2.$$
By using Property 5, we obtain that
\begin{align*}
&2\text{E}_{\text{all}}\{S_{\mathcal{P}}(G_1, G_2)S_{\mathcal{P}}(G_2, G_3)\}\\
&=\frac{2(n_1+n3)n_2}{N}+\frac{2(n_1+n_3)n_2(n_1+n_3+n_2-2)}{N(N-1)}\\
&~~~~+\frac{4(n_1+n_3)(n_1+n_3-1)n_2(n_2-1)}{N(N-1)}\\
&~~~~-\frac{2n_1n_2}{N}-\frac{2n_1n_2(n_1+n_2-2)}{N(N-1)}-\frac{4n_1(n_1-1)n_2(n_2-1)}{N(N-1)}\\
&~~~~-\frac{2n_3n_2}{N}-\frac{2n_3n_2(n_3+n_2-2)}{N(N-1)}-\frac{4n_3(n_3-1)n_2(n_2-1)}{N(N-1)}\\
&=\frac{4n_1n_3n_2(2n_2-1)}{N(N-1)}.
\end{align*}

To  prove Property 9, we again make use of the fact   that  the values of $p_{i,j}$ for cases (i) and (ii) are zeroes. We only need to consider the case (iii). This can be done by

  \begin{align*}
\text{E}_{\text{all}}\{S_{\mathcal{P}}(G_1, G_1)S_{\mathcal{P}}(G_2, G_2)\}&=(N-1)\times 0+2(N-2)\times0\\
 &+(N-2)(N-3)\frac{n_1(n_1-1)n_2(n_2-1)}{N(N-1)(N-2)(N-3)}.
\end{align*}
 
 \subsection{Proof of Theorem 3}
 
 We first consider a two-sample case for $k=2$ and $\bm{\theta}(F_{G_1})\neq \bm{\theta}(F_{G_2})$. By condition A4, the total weight  can be expressed as
\begin{align}\label{totd}
&\D^*(G_1, G_1)\sum_{i=1}^{N-1}I(v_i\in G_1, v_{i+1}\in G_1)+\D^*(G_2, G_2)\sum_{i=1}^{N-1}I(v_i\in G_2, v_{i+1}\in G_2)\nonumber\\
&+\D^*(G_1, G_2)\sum_{i=1}^{N-1}I\{(v_i\in G_1, v_{i+1}\in G_2)~\text{or}~(v_i\in G_2, v_{i+1}\in G_1)\}+\so_p(1).
\end{align}
Without loss of generality, we assume that $\D^*(G_1, G_1)<\min\{\D^*(G_1, G_2),\D^*(G_2, G_2)\}$ under condition A5.
Since we need to minimize the total costs, $\sum_{i=1}^{N-1}I(v_i\in G_1, v_{i+1}\in G_1)$ must be maximized. Therefore, the number of vertices belonging to $G_1$ and of degree 1 must be less than or equal to 2 in probability. Since the maximum vertex degree is limited to 2 and there are no cycles, 
$S_{\mathcal{P}^*}(G_1, G_2)\leq 2$ in probability.

Since the analysis of the total costs applies to the case of  $k$ samples with $k>2$,   $S_{\mathcal{P}^*}(g_{t_1}, g_{t_2})\leq  2$ in probability if $\bm{\theta}(F_{g_{t_1}})\neq \bm{\theta}(F_{g_{t_2}})$. Theorem 3 follows from this.

\subsection{Proof of Theorem 4}
Under the null hypothesis,
$$\text{E}_{\text{all}}\{S_{\mathcal{P}}(G_m, G_\ell)\}=O(N)~\text{and}~ \text{Var}_{\text{all}}\{S_{\mathcal{P}}(G_m, G_\ell)\}=O(N).$$ Therefore, $w_{i,j}=O(N^{-1/2})$, and 
 $N^{-1/2}\bar Z_{\mathcal{P}}=c_1+\so_{p}(1)$ for $c_1>0$. 
 
   By Theorem 3, under the alternative hypothesis, $N^{-1/2}\bar Z_{\mathcal{P}}=c_2+\so_{p}(1)$ with   $0<c_2<c_1$. Hence, the power based on $\bar{Z}_{\mathcal{P}}$ converges to 1.

For the minimum statistic, under the alternative hypothesis,  we have 
 $$P(\underline{Z}_{\mathcal{P}^*}\leq x)\geq P\left[w_{i,j}\left\{S_{\mathcal{P}^*}(G_m, G_\ell)-\text{E}_{\text{all}}S_{\mathcal{P}}(G_m, G_\ell)\right\}\leq x\right].$$
 
As $S_{\mathcal{P}^*}(G_m, G_\ell)\leq 2$ in probability and  $w_{i,j}\text{E}_{\text{all}}S_{\mathcal{P}}(G_m, G_\ell)\rightarrow\infty$, the right probability converges to 1. Therefore, $P(\underline{Z}_{\mathcal{P}^*}\leq x)$ converges to 1. The proof is finished.

\begin{thebibliography}{7}
\expandafter\ifx\csname natexlab\endcsname\relax\def\natexlab#1{#1}\fi







  	\bibitem[{Shalon et~al. (1996)}]  {SSB96}   
  	\textsc{Shalon, D.}, \textsc{Smith, S. J.} \& \textsc{Brown, P. O.}
  	(1996). 
  	\newblock{A DNA microarray system for analyzing complexDNA samples using two-color fluorescent probehybridization}. 
  	\newblock \textit{Genome Research} \textbf{6}, 639–645.


  	\bibitem[{Weinstein et~al. (2013)}]  {We13}   
\textsc{Weinstein, J. N.}, \textsc{Collisson, E. A.}, \textsc{Mills, G. B.}, \textsc{Shaw, K. R.}, \textsc{Ozenberger, B. A.}, \textsc{Ellrott, K.}, \textsc{Shmulevich, I.}, \textsc{Sander, C.} \& \textsc{Stuart, J. M.} (2013).
  	\newblock{The Cancer Genome Atlas Pan-Cancer analysis project}. 
  	\newblock \textit{Nature Genetics} \textbf{45}, 1113–1120.




  	\bibitem[{Zhou and Rossi (2017)}]  {ZR17}   
  	\textsc{Zhou, J.}  \& \textsc{Rossi, J.}
  	(2017). 
  	\newblock{Aptamers as targeted therapeutics: current potential and challenges}. 
  	\newblock \textit{Nature Reviews Drug Discovery } \textbf{16}, 181–202.
 
 
  	\bibitem[{Lehallier et~al. (2019)}]  {Le19}   
 \textsc{Lehallier, B.}, \textsc{Gate, D.}, \textsc{Schaum, N.}, \textsc{Nanasi, T.}, \textsc{Lee, S. E.}, \textsc{Yousef, H.}, \textsc{Losada, P. M.}, \textsc{Berdnik, D.}, \textsc{Keller, A.}, \textsc{Verghese, J.}, \textsc{Sathyan, S.}, \textsc{Franceschi, C.}, \textsc{Milman, S.}, \textsc{Barzilai, N.} \& \textsc{Wyss-Coray, T.}  (2019).
  	\newblock{Undulating changes in human plasma proteome profiles across the lifespan}. 
  	\newblock \textit{Nature Medicine} \textbf{25}, 1843–1850.



\bibitem[{Jung et~al. (2019)}]  {JSK19}   
  	\textsc{Jung, W. B.}, \textsc{Shim, H. J.} \& \textsc{Kim, S. G.}
  	(2019). 
  	\newblock{Mouse BOLD fMRI at ultrahigh
field detects somatosensory networks including thalamic nuclei}. 
  	\newblock \textit{NeuroImage} \textbf{195}, 203–214.


 
  	\bibitem[{Bai and Saranadasa (1996)}]  {BS96}   
  	\textsc{Bai, Z.}  \& \textsc{Saranadasa, H.}
  	(1996). 
  	\newblock{Effect of high dimension: by an example of a two sample problem}. 
  	\newblock \textit{Statistica Sinica} \textbf{6}, 311–329.

	\bibitem[{Chen and Qin (2010)}]  {CQ10}   
  	\textsc{Chen, S.}  \& \textsc{Qin, Y.}
  	(2010). 
  	\newblock{A two-sample test for high-dimensional data with applications to gene-set testing}. 
  	\newblock \textit{Annals of Statistics} \textbf{38}, 808–835.
 

\bibitem[{Cai et~al. (2014)}]  {CLX14}   
  	\textsc{Cai, T.}, \textsc{Liu, W.} \& \textsc{Xia, Y.}
  	(2014). 
  	\newblock{Two-sample test of high dimensional means under dependence}. 
  	\newblock \textit{Journal of the Royal Statistical Society: Series B} \textbf{76}, 349–372.


\bibitem[{Xu et~al. (2016)}]  {XLWP16}   
  	\textsc{Xu, G.}, \textsc{Lin, L.}, \textsc{Wei, P.} \& \textsc{Pan, W.}
  	(2016). 
  	\newblock{An adaptive two-sample test for high-dimensional means}. 
  	\newblock \textit{Biometrika} \textbf{103}, 609–624.


 \bibitem[{He et~al. (2021)}]  {HXWP21}   
  	\textsc{He, Y.}, \textsc{Xu, G.}, \textsc{Wu, C.} \& \textsc{Pan, W.}
  	(2021). 
  	\newblock{Asymptotically independent U-statistics in high-dimensional testing}. 
  	\newblock \textit{Annals of Statistics} \textbf{49}, 154-181.

 
 
	\bibitem[{Srivastava and Kubokawa (2013)}]  {SK13}   
  	\textsc{Srivastava, M. S.}  \& \textsc{Kubokawa, T.}
  	(2013). 
  	\newblock{Tests for multivariate analysis of variance in high dimension under non-normality}. 
  	\newblock \textit{Journal of Multivariate Analysis} \textbf{115}, 204–216.


 \bibitem[{Hu et~al. (2017)}]  {HBWW17}   
  	\textsc{Hu, J.}, \textsc{Bai, Z.}, \textsc{Wang, C.} \& \textsc{Pan, W.}
  	(2017). 
  	\newblock{On testing the equality of high dimensional mean vectors with unequal covariance matrices}. 
  	\newblock \textit{Annals of the Institute of Statistical Mathematics} \textbf{69}, 365–387.

 
 	\bibitem[{Friedman and Rafsky (1979)}]  {FR79}   
  	\textsc{Friedman, J. H.}  \& \textsc{Rafsky, L. C.}
  	(1979). 
  	\newblock{Multivariate generalizations of the Wald-Wolfowitz and Smirnov two-sample tests}. 
  	\newblock \textit{Annals of Statistics} \textbf{7}, 697–717.


 \bibitem[{Biswas et~al. (2014)}]  {BMG14}   
  	\textsc{Biswas, M.}, \textsc{Mukhopadhyay, M.}  \& \textsc{Ghosh, A. K.}
  	(2014). 
  	\newblock{A distribution-free two-sample run test applicable to high-dimensional data}. 
  	\newblock \textit{Biometrika} \textbf{101}, 913–926.

  	\bibitem[{Wald and Wolfowitz (1940)}]  {WW40}   
  	\textsc{Wald, A.}  \& \textsc{Wolfowitz, J.}
  	(1940). 
  	\newblock{On a test whether two samples are from the same distribution}. 
  	\newblock \textit{Annals of Mathematical Statistics} \textbf{11}, 147–162.
 
   	\bibitem[{Chen and Zhang (2013)}]  {CZ13}   
  	\textsc{Chen, H.}  \& \textsc{Zhang, N. R.}
  	(2013). 
  	\newblock{Graph-based tests for two-sample comparisons of categorical data}. 
  	\newblock \textit{Statistica Sinica} \textbf{23}, 1479-1503.
 

 \bibitem[{Chen et~al. (2018)}]  {CCS18}   
  	\textsc{Chen, H.}, \textsc{Chen, X.}  \& \textsc{Su, Y.}
  	(2018). 
  	\newblock{A weighted edge-count two sample test for multivariate and object data}. 
  	\newblock \textit{Journal of the American Statistical Association: Theory and Methods} \textbf{113}, 1146-1155.

 \bibitem[{Chen and Friedman (2017)}]  {CF17}   
  	\textsc{Chen, H.}  \& \textsc{Friedman, J. H.}
  	(2017). 
  	\newblock{A new graph-based two-sample test for multivariate and object data}. 
  	\newblock \textit{Journal of the American Statistical Association: Theory and Methods} \textbf{112}, 397-409.
 

 \bibitem[{Zhang and Chen (2021)}]  {ZC21}   
  	\textsc{Zhang, J.}  \& \textsc{Chen, H.}
  	(2021). 
  	\newblock{Graph-based two-sample tests for data with repeated observations}. 
  	\newblock \textit{Statistica Sinica}, to appear.
 
 

 \bibitem[{Mukhopadhyay and Wang (2020)}]  {MW20}   
  	\textsc{WMukhopadhyay, S.}  \& \textsc{Wang, K.}
  	(2020). 
  	\newblock{A nonparametric approach to high-dimensional k-sample comparison problems}. 
  	\newblock \textit{Biometrika} \textbf{107}, 555–572.
 

   \bibitem[{Genz (1992)}]  {Ge92}   
  	\textsc{Genz, A.}
  	(1992). 
  	\newblock{Numerical computation of multivariate normal probabilities}. 
  	\newblock \textit{Journal of Computational and Graphical Statistics} \textbf{1}, 141-150 .
 
   \bibitem[{Sz\'ekely  and Rizzo (2005)}]  {SR05}   
  	\textsc{Sz\'ekely, G. J.}  \& \textsc{Rizzo, M. L.}
  	(2020). 
  	\newblock{Hierarchical clustering via joint between-within distances:  extending Ward’s minimum variance method}. 
  	\newblock \textit{Journal  of  Classification} \textbf{22}, 151–183.
 

\bibitem[{Benjamini and Hochberg (1995)}]{BH95}   
  	\textsc{Benjamini, Y.} \& \textsc{Hochberg, Y.}
  	(2014). 
  	\newblock{Controlling the false discovery rate: a practical and powerful approach to multiple testing}. 
  	\newblock \textit{Journal of the Royal Statistical Society: Series B} \textbf{57}, 289–300.
 
 



  

   



\end{thebibliography}
\end{document}